\documentclass{cta-author}

{}
{}
{}

\usepackage{breqn}
\usepackage{adjustbox}
\DeclareUnicodeCharacter{FF0C}{\textvisiblespace}

\usepackage[english]{babel}
\usepackage[utf8]{inputenc}
\usepackage{algorithm}
\usepackage[noend]{algpseudocode}
\usepackage{amsfonts}

\usepackage{listings}
\usepackage{fixltx2e}
\usepackage{subfigure}
\usepackage{amsmath}
\usepackage[
	n,
	operators,
	advantage,
	sets,
	adversary,
	landau,
	probability,
	notions,	
	logic,
	ff,
	mm,
	primitives,
	events,
	complexity,
	asymptotics,
	keys]{cryptocode}
\usepackage{tabto}
\usepackage{array}

\begin{document}

\title{Incentive-weighted Anomaly Detection for False Data Injection Attacks Against Smart Meter Load Profiles}

\author{\au{Martin Higgins\textsuperscript{1*}, }
\au{Bruce Stephen\textsuperscript{2}, }
and 
\au{David Wallom\textsuperscript{1}}
}

\address{\add{1}{Oxford e-Research Center, Department of Engineering Science, University of Oxford, Holywell House, 64 Osney Mead, Oxford, OX2 0ES, United Kingdom}
\add{2}{Advanced Electrical Systems Research
Group, Institute of Energy and Environment, University of Strathclyde,
Glasgow G1 1RD}
\email{martin.higgins@eng.ox.ac.uk}}

\begin{abstract}
Spot pricing is often suggested as a method of increasing demand-side flexibility in electrical power load. However, few works have considered the vulnerability of spot pricing to financial fraud via false data injection (FDI) style attacks. In this paper, we consider attacks which aim to alter the consumer load profile to exploit intraday price dips. We examine an anomaly detection protocol for cyber-attacks that seek to leverage spot prices for financial gain. In this way we outline a methodology for detecting attacks on industrial load smart meters. We first create a feature clustering model of the underlying business, segregated by business type. We then use these clusters to create an incentive-weighted anomaly detection protocol for false data attacks against load profiles. This clustering-based methodology incorporates both the load profile and spot pricing considerations for the detection of injected load profiles. To reduce false positives, we model incentive-based detection, which includes knowledge of spot prices, into the anomaly tracking, enabling the methodology to account for changes in the load profile which are unlikely to be attacks.
\end{abstract}

\maketitle

\section{Introduction}
The contemporary power network is a cyber-physical system consisting of modern communications technologies working in conjunction with sophisticated power electronics. Up until recently, most of the power system innovations in real-time monitoring occurred at the transmission layer. However, the recent introduction of smart metering offers exciting opportunities for distribution level consumers and system operators to optimise their consumption of power. The increased granularity offered by load profile data offers new ways to reduce costs and encourage demand-side flexibility \cite{Gelazanskas2014DemandDirection} \cite{Aduda2016DemandImplications}. Increasingly, variable tariffs are becoming popular which offer intraday variation in the electricity consumption price. These types tariffs make utility level spot prices directly available to industrial consumers themselves \cite{Garcia1985ThePricing}. Since consumers can receive cheaper prices to consume power during non-peak hours, they can save cash if they act to adjust their demand curves. Advantages can also extend to the system operators in terms of potential benefits from lower intraday volatility in consumption. However, risks have emerged along which are not currently considered in the present frameworks. On the one hand, smart meters can enable consumers to use spot pricing, which unlocks rewards for proactive consumers; on the other hand, the large intraday volatility of spot pricing means there is a direct cash incentive for malicious actors with the capabilities to bypass the relatively basic smart-metering cyber-infrastructure. These cash incentives are amplified for industrial users for whom electricity demand far outstrips the average consumer. In view of this, it is necessary to begin considering how users may try and exploit the system. 

\subsection{Categorisation of Load Profiles}
With the growing ubiquity of smart metering, researchers are increasingly investigating how to effectively utilise the data they capture. Load profile categorisation, either via clustering or using other techniques, has also become a popular sub-field. Almost all studies involving smart meter load profiles have focused on residential smart meter data. For example, the authors in \cite{Kwac2014HouseholdData} explored a segmentation strategy for households using hourly data, while a clustering approach for consumer smart meter data was examined in \cite{McLoughlin2015AData}. As further examples, prior research works have identified behavioural demand profiles using smart meter data \cite{Haben2016AnalysisData}, used C-Vine Copula models for capturing intra-day variability  \cite{Sun2017C-VineData}, used non-gaussian residual to model intra-day forcasting at the feeder level \cite{Bruce2020Non-GaussianFeeders},  explored novel approaches for load profiling using smart meter data \cite{Khan2018AData}, and used load profiles to cluster consumer profiles \cite{Tureczek2018ElectricityData}. In \cite{Stephen2014EnhancedCustomers} Stephen \textit{et al} presented several Linear Gaussian (LG) load profiling techniques. These were embedded within a mixture model framework, which allowed multiple behaviours to be considered with the most probable used for categorization. The prior focus on consumer data is likely attributable to the relative availability of this data in comparison to industrial load flow profiles. By contrast, some works have addressed non-residential flows, including the authors in \cite{Verdu2006ClassificationMaps}, which used self-organizing maps to classify industrial loads. Previous works have also used customer-specific data to create use profiles u\cite{Rasanen2010Data-basedData} and analysed industrial electricity consumption with respective to behavioural dynamics \cite{Wang2016ClusteringApplications}. The authors in \cite{Chicco2012OverviewGrouping} introduced a general scheme for analyzing load patterns, while an overview of clustering techniques was presented in \cite{Chicco2004LoadCustomers}, which summarises and evaluates methods for load pattern classification. Often, these works stop short of finding a use case for the profiling. 
In \cite{Zhan2020BuildingBenchmarking}, the authors applied a clustering-based framework for building energy-based benchmarks. Data extracted from smart meter load profiles were used to categorise buildings according to their operational characteristics. In \cite{Granell2021PredictingLearning}, the load profiles of supermarket chains were predicted using machine learning. In \cite{Elnozahy2013AAlgorithms} a novel probabilistic approach was proposed that utilises similar principal components. Hu \textit{et al}. \cite{Hu2021ClassificationClustering} used interpretable feature extraction to categorize load profiles based on a combination of statistical and temporal features. The authors in \cite{Wang2015LoadReview} examined load profiling and its applications in relation to demand response. An anomaly detection scheme for big industrial datasets is applied in \cite{Caithness2018AnomalyData}. A review of electric load classification in smart grids is available in \cite{Zhou2013AEnvironment}.

\subsection{Attacks Against Metering Infrastructure}
Before the use of smart meter infrastructure, bypassing an electricity meter was a common method of defrauding utility operators. However, from the perspective of utility providers, direct bypass attacks are easy to identify using data driven methods as they are effectively a string of 0s. In the case of smart meter infrastructure, while some commentators initially believed that smart meters would provide additional security, they have been proven to susceptible to hacking \cite{Tangsunantham2013ExperimentalMeters}. 

The available evidence suggests that in the future, smart meter attacks may aim to change the transmitted load curve completely, thereby reducing the cost of power consumption. In the past, these types of attacks have been called False Data Injection (FDI) attacks and have usually been suggested at the transmission layer. A review of this attack type can be found in \cite{Wang2019ReviewSystem}. We explored these type of attacks previously in  \cite{Higgins2021EnhancedDefence}, \cite{Higgins2022Cyber-PhysicalDefences} \& \cite{Higgins2021TopologyInformation} in case studies where FDI attackers alter system measurements to spoof the transmission-level state estimation processes. However, while FDI style attacks on transmission level infrastructure have received significant research attention, limited research has examined the impact of these attacks on distribution-level systems such as smart meter load profiles. A putative advantage of the FDI approach is that these attacks can utilise distributed, poorly protected measurements rather than attacking a well-defended central system operator. This is especially true for distribution-level attacks against metering infrastructure as these devices are usually decoupled from operational processes and not monitored in real-time by utility providers. Indeed, modern smart meter infrastructure has also been shown to suffer from several vulnerabilities, which can be exploited by motivated attackers \cite{Ur-Rehman2015SecuritySystems}. 

\subsection{Contributions}
While many papers have addressed the categorisation or clustering of load profile data, few demonstrate the utility, or action that results from this categorisation. In this work, we propose both a methodology for grouping load profile data and also an application for this process within the realms of cyber-attacks. 

The main contributions of this work are as follows:

\begin{itemize}
    \item This work introduces a new methodology for the clustering of load profile data. This methodology involves a two-step process that incorporates both clustering and silhouette scoring to establish a set of base models within each industry type. We use 20 features for this approach, which include a combination of global statistical, index and quartile statistical features. 
    \item We use these average cluster groups to produce a scoring model for new inbound datasets. This scoring model incorporates both model departure and spot prices to present an incentive-weighted model of fraud detection in load profiles. This incentive weighted approaches considers not simply departure from the expected model but also the weighted spot price to identify when attackers maybe trying to change profiles for financial gain. 
    \item Finally, we develop our model, using real load profile data and real-life spot price data and test it using simulated FDI attacks on load profiles. This evaluates the effectiveness of the detection approach in the face of sophisticated cyber-attacks. 
\end{itemize}

The next section presents the base model methodology used to build the average cluster models.

\section{Base Model Methodology}

\subsection{Input Data}
The input data were obtained as part of the Energy Demand Research Project (EDRP). The EDRP aims to understand and model how load user flexibility changes as consumers develop an awareness and understanding of their energy consumption. The dataset consists of industrial load flow profiles for 12,055 businesses operating over a two-year period. Within these businesses, we categorized business data, and took a subset of the businesses under the branch of consumer entertainment industrial parks. The reason we opted for this subset is that these businesses offer distinct business models that are easily interpretable, at a conceptual level, to the average user. The profiles consist of 48 consumption periods, with each period corresponding to 30-minute power consumption windows within a given 24 hour day. Within this, we focus on summer profile data sets (June through September) to maintain consistency in the underlying data. 

\subsection{Data Pre-processing}
We consider that while some businesses may share similar relative properties in overall power consumption, the magnitude of energy consumption within business of the same type may vary considerably. When building our groupings, we intend to identify businesses based on the shape of operation and relative properties rather than straight magnitude. Therefore, for each individual business, we perform a max-min normalization of the load profile data using the following equation:

\begin{equation}
    z_i = \frac{x_i- min(x)}{max(x)-min(x)}
    \label{minmax}
\end{equation}

In this max-min normalisation equation, $x=(x_1,\ldots,x_n)$ represents an array of length equal to the number of load consumption measurements for a given business line. The normalisation enables us to capture departures from expected operation demand curves. Simple changes in consumption magnitude (such as a bypass attack) are typically easy to identify via conventional means, and so we focus on relative model departure. We also apply a `low touch' data cleaning strategy, which aims to remove corrupted or incomplete datasets to the greatest possible extent, while minimising the discarded data. It is often tempting to be overzealous when cleaning data, but as we are working with a real-world data sample, we wish to reduce the amount of data discarded.

\subsection{Feature-based Clustering}
We use feature-based clustering to establish the base models for our anomaly detection and incentive-weighted anomaly detection system. We use a set of 20 different features consisting of global statistical features, quartile statistical features, and index-based features. Table \ref{table1} summarises the features used in our clustering algorithm. The approach employed is similar to the one outlined in \cite{Hu2021ClassificationClustering}, with the exception that we also incorporate quartile statistical values.

\begin{table}[]
\begin{tabular}{|l|l|l|}
\hline
\textbf{Feature   No.} & \textbf{Feature   Description} & \textbf{Feature Type} \\ \hline
G1                     & Mean                           & Global                \\ \hline
G2                     & Standard Deviation             & Global                \\ \hline
G3                     & Max                            & Global                \\ \hline
G4                     & Min                            & Global                \\ \hline
G5                     & Range                          & Global                \\ \hline
G6                     & Sum                            & Global                \\ \hline
G7                     & Skew                           & Global                \\ \hline
G8                     & Kurtosis                       & Global                \\ \hline
Q9                     & Sum 1-12                       & Quartile              \\ \hline
Q10                    & Sum 12-24                      & Quartile              \\ \hline
Q11                    & Sum 24-36                      & Quartile              \\ \hline
Q12                    & Sum 36-48                      & Quartile              \\ \hline
Q13                    & Standard Deviation   1-12      & Quartile              \\ \hline
Q14                    & Standard Deviation   12-24     & Quartile              \\ \hline
Q15                    & Standard Deviation   24-36     & Quartile              \\ \hline
Q16                    & Standard Deviation   36-48     & Quartile              \\ \hline
I17                    & Max Time Period                & Index                 \\ \hline
I18                    & Min Time Period                & Index                 \\ \hline
I19                    & Index \textgreater{}Mean       & Index                 \\ \hline
I20                    & Index \textless{}Mean          & Index                 \\ \hline
\end{tabular}
\caption{List of features used in clustering model.}
\label{table1}
\end{table}

\subsection{Hierarchical Clustering}
This subsection outlines the proposed combination of clustering and scoring used to establish the average profile groupings and cluster numbers. After data pre-processing, we perform agglomerative hierarchical clustering on the respective industrial load business types. Hierarchical clustering is also known as AGNES (agglomerative nesting) and refers to a bottom-up approach to clustering wherein each observation starts in a cluster on its own and clusters are slowly merged. The steps involved in AGNES are as follows:

\begin{enumerate}
    \item The proximity matrix for each point within the dataset is calculated.
    \item The algorithm then considers each element as a cluster consisting of a single element cluster.
    \item The two closest clusters are merged and the new proximity matrix is recalculated for the dataset.
    \item Steps 1-3 are then repeated until the desired number of clusters is reached. 
\end{enumerate}

As we are using an unsupervised learning approach, we sought to avoid manually inputting a cluster number for the algorithm to use. Therefore, we incorporate an automatic cluster number selection feature, which utilizes silhouette scoring. 

\subsection{Silhouette Coefficient}
The silhouette coefficient is a method of quality checking and validating cluster consistency within groups. The coefficient measures the similarity of an object with respect to its given cluster. Each data point within a series is assigned a silhouette value. This silhouette value of an individual data point is given by 

\begin{equation}
    s(\textbf{z}) = \frac{b(z)-a(z)}{\max(a(z),b(z))}, 
    \label{minmax}
\end{equation}

where $s(z)$ is the silhouette score for a given data point $z$, and $b(z)$ is the average minimum distance between $z$ and the clusters that $z$ is not located within and $a(z)$ is the average distance between $z$ and all the other data points with the cluster $z$ is located within. 

The silhouette coefficient is then given by finding the maximum value of the mean silhouette score for a given number of clusters $k$ such that 

\begin{equation}
    SC = \max_{k} \overline{s}(k),
    \label{minmax}
\end{equation}

where $\overline{s}(k)$ is the mean silhouette score across the entire dataset. 

In this work, we employ a short loop. This compares the silhouette coefficient under different cluster numbers (up to 5) and selects for the maximum coefficient value. In turn, this is used to define the number of clusters involved in hierarchical clustering.

\section{Incentive-based Anomaly Tracking}
Historically, anomaly tracking in energy systems has been based on the analysis of simple departures from expected models. However, within the context of cyber-attacks, model departure is not necessarily an indication of foul play. We consider that departure from a model is not merely an indication of a cyber-attack. Anomalous measurements do occur amid routine operation. We also consider that in a scenario considering financial or cost lowering attack, the attacker is unlikely to inject an attack vector which will increase his overall cost. Therefore, we can use considerations about the attack vector incentive as a method of reducing false positives. In this way we create an 'incentive-based' anomaly detection which considers the cash incentive of the attack as well as the direct anomaly.   

\subsection{Detection Model}

Here we outline the detection model for the incentive-weighted anomaly detection. The steps involved are as follows:

\begin{enumerate}
    \item The outlined hierarchical and silhouette scoring clustering model are leveraged to model expected behaviours in load profiles for respective industry types.
    \item Unexpected departures from the underlying models in new inbound data are identified for the respective company groups. Also, a score is created based on how different these groups are, which is referred to as the violation scoring.
    \item We then use the weighted average spot price to produce an incentive based scoring model which indicates whether a profile is financially preferable. 
    \item The scores are then combined to establish the incentive-weighted violation score to identify potential FDI attacks based on model difference and potential financial gain.
\end{enumerate}

\subsection{Violation Score}
We consider a metric, dubbed a violation score, used to assess how different a new incoming dataset is compared to the previous model. The violation score is based on how often these inbound measurements violate a confidence interval of 2 standard deviations when compared to the average model for the business group. We start with the following equation:

\begin{equation}
    \textbf{VSD}^n = \textbf{2ASD}-||(\textbf{z}_{new}^n - \textbf{AC})||,
    \label{Violationscore}
\end{equation}

where $\textbf{VSD}^{p}$ is an array of length $t$ that contains the respective violation decisions for a given consumption interval, $\textbf{AC}$ is an array of length $t$ representing the average cluster profile, $\textbf{ASD}$ is an array of length $t$ representing the standard deviations for the respective consumption periods, and $\textbf{z}_{new}$ is an array of length $t$ that represents the new measurement set which is being checked. Also, $n$ refers to the number of days in the set. A violation is recorded if $VS$ is a negative value such that

\begin{equation}
   VSC_{t}^n =
    \begin{cases}
      1 & \text{if $VSD_t^n < 0$}\\
      0 & \text{if $VSD_t^n > 0$}
    \end{cases}       
    \label{Violationscore2}
\end{equation}

In turn, this is presented as a percentage of the number of periods recorded:

\begin{equation}
    VSP = \sum_{n=1}^n\sum_{t=1}^t (\frac{VSC_{t}^n}{nt}),
    \label{Violationscore}
\end{equation}

where $VSP$ is the violation score. This score gives an initial indication as to whether there is a significant departure from the previously established cluster groups. A high violation score indicates that the model varies significantly from the average cluster model established by the clustering algorithm. We consider that a simple departure from the underlying model is not necessarily an indication of foul play and that we must also consider the impact of an attack. 

\subsection{Incentive Score}

We consider incentive as a product of the relative gain that a change from the average profile gives to a customer. To do this, we incorporate the weighted average spot price $WSP$ versus flat price to identify regions where there may be an incentive to change the input profile. The weighted spot price array is calculated as below:

\begin{equation}
    \textbf{WSP} = \frac{(\textbf{CSP} - \textbf{FP})}{FP},
    \label{weightedspotprice}
\end{equation}

where $\textbf{WSP}$ is an array of length $t$ that represents the number of prices in the period (in this case, 48 half-hourly periods), and $\textbf{FP}$ is an array of length $t$ consisting of the flat price $FP$. The weighted spot price is then used to score the incentives given by the departure from the model, such that

\begin{equation}
    ISC = ||\textbf{z}_{new}^n-\textbf{AVCM} \cdot \textbf{WSP}||,
    \label{incentivescore}
\end{equation}

where $ISC$ is the incentive score, $\textbf{z}_{new}$ is an array of normalised load profile measurements of length $t$, and $\textbf{AVCM}$ is the given average cluster profile. This gives an initial indication as to whether there is a significant departure from the underlying model. A high violation score indicates that the model varies significantly from the average cluster model established by the clustering algorithm.

\subsection{Incentive-weighted Violation Score}
Finally, we consider the weighted incentive-based violation score $WIVS$ as a simple product of the violation score and the incentive score such that  

\begin{equation}
    WIVS =  ISC \cdot VSP
    \label{weightedincentivescore}
\end{equation}

This yields a simple metric for each given business, with which it is possible to assess the likelihood of financial fraud. In the following section, these metrics are tested with existing business types and FDI profile sets to verify the effectiveness of the approach.

\section{Results}
In this section we outline the results for the clustering technique, average profiles and the incentive-weighted detection algorithm. We initially show the results of the average profile model and then go onto examine the results of the anomaly detection algorithm on a combination of future load profile data and injected red team profiles. 
\begin{figure*}
     \centering
     \begin{subfigure}
         \centering
         \includegraphics[width=2.4in]{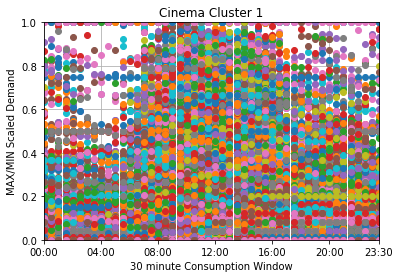}
      \end{subfigure}
     \begin{subfigure}
         \centering
         \includegraphics[width=2.4in]{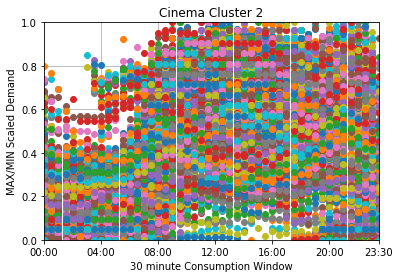}
         \end{subfigure}
     \begin{subfigure}
         \centering
         \includegraphics[width=2.4in]{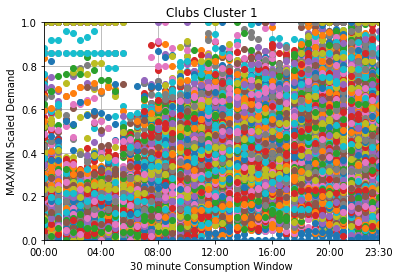}
         \end{subfigure}    
     \begin{subfigure}
         \centering
         \includegraphics[width=2.4in]{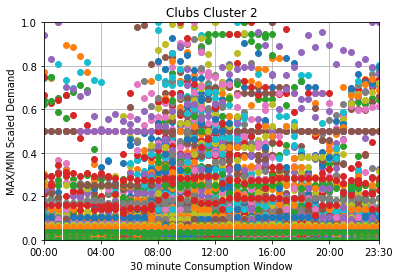}
         \end{subfigure}    
     \begin{subfigure}
         \centering
         \includegraphics[width=2.4in]{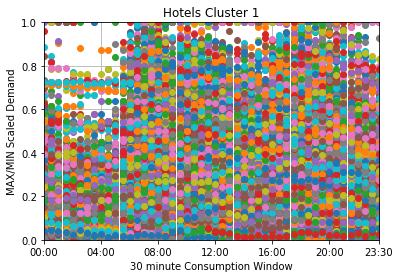}
         \end{subfigure}   
     \begin{subfigure}
         \centering
         \includegraphics[width=2.4in]{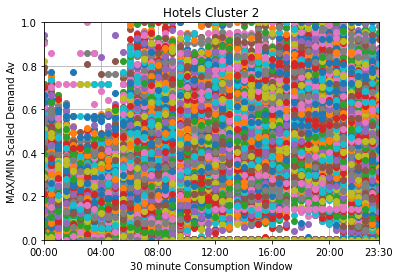}
         \end{subfigure}
     \begin{subfigure}
         \centering
         \includegraphics[width=2.4in]{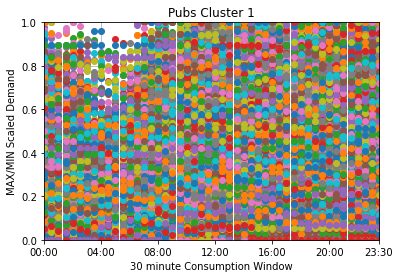}
         \end{subfigure}    
     \begin{subfigure}
         \centering
         \includegraphics[width=2.4in]{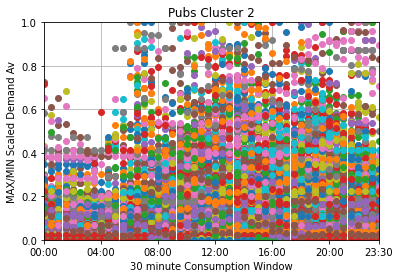}
         \end{subfigure}            
        \caption{All data points for respective industry clusters. Based on weekend profile data during summer 2009.}
        \label{clustergroupings}
\end{figure*}

\begin{figure*}
     \centering
     \begin{subfigure}
         \centering
         \includegraphics[width=2.4in]{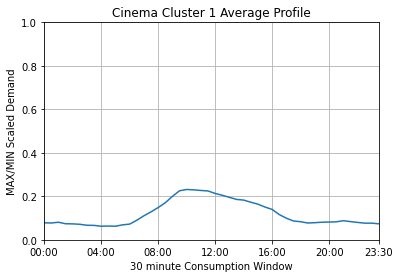}
      \end{subfigure}
     \begin{subfigure}
         \centering
         \includegraphics[width=2.4in]{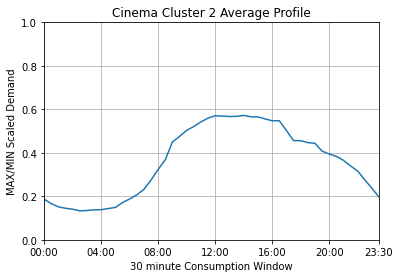}
         \end{subfigure}
     \begin{subfigure}
         \centering
         \includegraphics[width=2.4in]{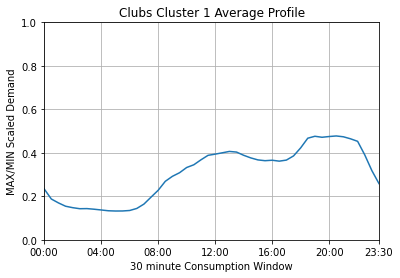}
         \end{subfigure}    
     \begin{subfigure}
         \centering
         \includegraphics[width=2.4in]{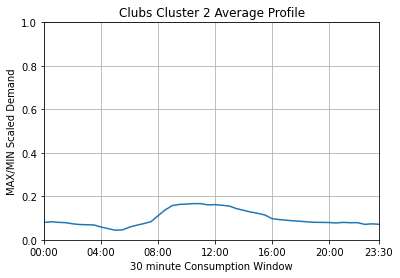}
         \end{subfigure}    
     \begin{subfigure}
         \centering
         \includegraphics[width=2.4in]{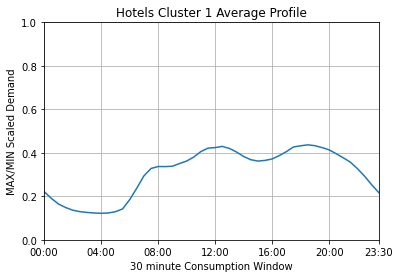}
         \end{subfigure}   
     \begin{subfigure}
         \centering
         \includegraphics[width=2.4in]{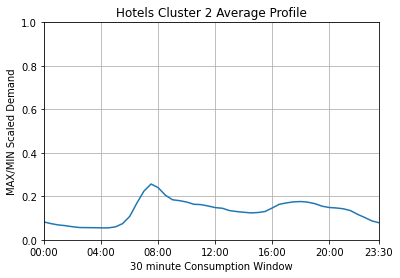}
         \end{subfigure}
     \begin{subfigure}
         \centering
         \includegraphics[width=2.4in]{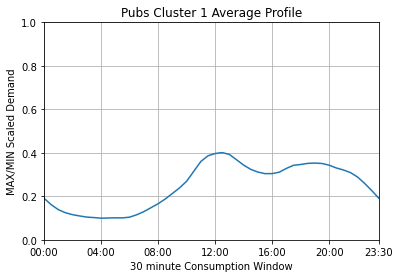}
         \end{subfigure}    
     \begin{subfigure}
         \centering
         \includegraphics[width=2.4in]{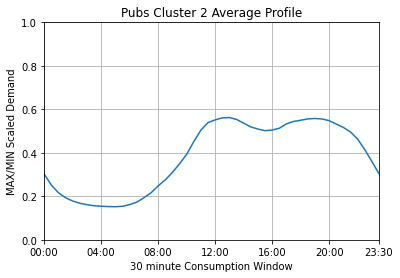}
         \end{subfigure}            
        \caption{Average cluster based on weekend profile data during summer 2009.}
        \label{averageclustergroupings}
\end{figure*}

\begin{figure*}
     \centering
     \begin{subfigure}
         \centering
         \includegraphics[width=2.4in, height = 2in]{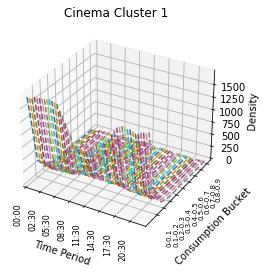}
      \end{subfigure}
     \begin{subfigure}
         \centering
         \includegraphics[width=2.4in, height = 2in]{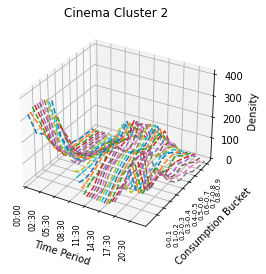}
         \end{subfigure}
     \begin{subfigure}
         \centering
         \includegraphics[width=2.4in, height = 2in]{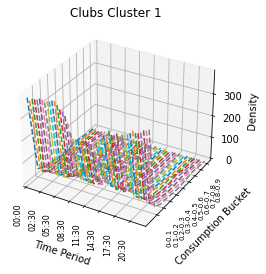}
         \end{subfigure}    
     \begin{subfigure}
         \centering
         \includegraphics[width=2.4in, height = 2in]{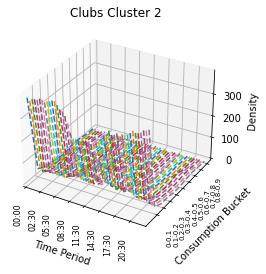}
         \end{subfigure}    
     \begin{subfigure}
         \centering
         \includegraphics[width=2.4in, height = 2in]{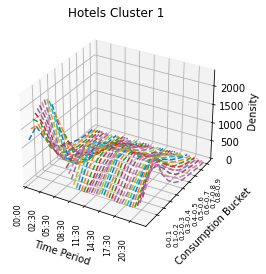}
         \end{subfigure}   
     \begin{subfigure}
         \centering
         \includegraphics[width=2.4in, height = 2in]{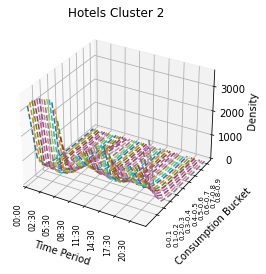}
         \end{subfigure}
     \begin{subfigure}
         \centering
         \includegraphics[width=2.4in, height = 2in]{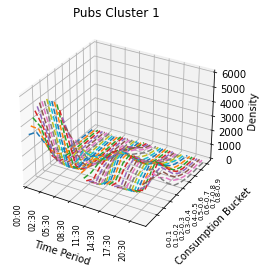}
         \end{subfigure}    
     \begin{subfigure}
         \centering
         \includegraphics[width=2.3in, height = 2in]{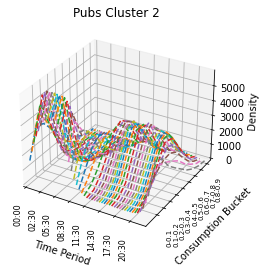}
         \end{subfigure}            
        \caption{3d histograms of all data points for respective industry clusters. Based on weekend profile data during summer 2009.}
        \label{Histograms}
\end{figure*}

\begin{figure*}
     \centering
     \begin{subfigure}
         \centering
         \includegraphics[width=1.6in]{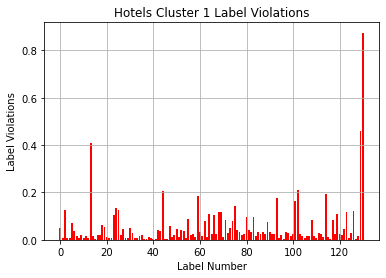}
      \end{subfigure}
     \begin{subfigure}
         \centering
         \includegraphics[width=1.6in]{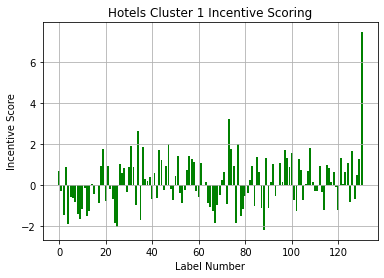}
         \end{subfigure}
     \begin{subfigure}
         \centering
         \includegraphics[width=1.6in]{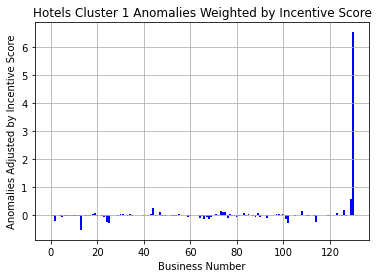}
         \end{subfigure}

     \begin{subfigure}
         \centering
         \includegraphics[width=1.6in]{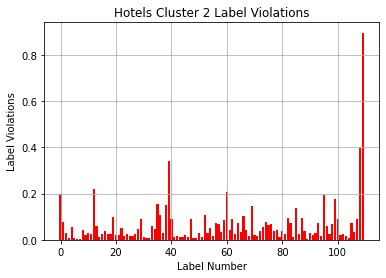}
      \end{subfigure}
     \begin{subfigure}
         \centering
         \includegraphics[width=1.6in]{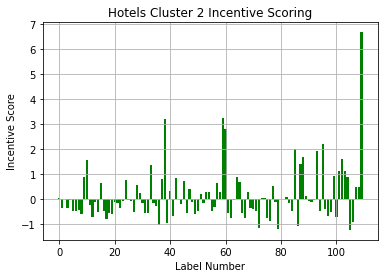}
         \end{subfigure}
     \begin{subfigure}
         \centering
         \includegraphics[width=1.6in]{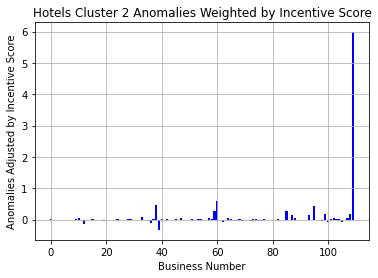}
         \end{subfigure} 
        \caption{Hotel label violations, incentive score, and weighted incentive score.}
        \label{hotelviolations}
\end{figure*}

\begin{figure*}
     \centering
     \begin{subfigure}
         \centering
         \includegraphics[width=1.6in]{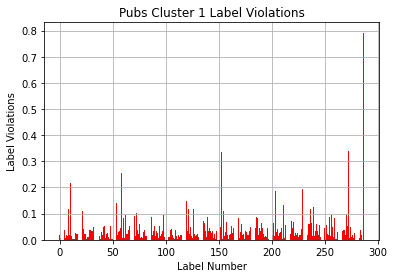}
      \end{subfigure}
     \begin{subfigure}
         \centering
         \includegraphics[width=1.6in]{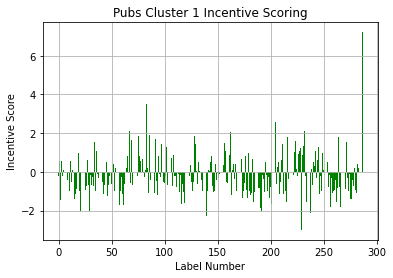}
         \end{subfigure}
     \begin{subfigure}
         \centering
         \includegraphics[width=1.6in]{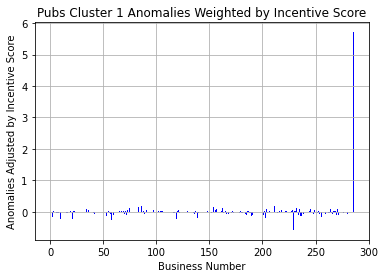}
         \end{subfigure}

     \begin{subfigure}
         \centering
         \includegraphics[width=1.6in]{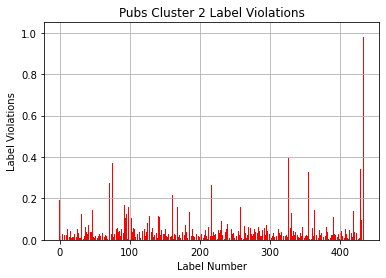}
      \end{subfigure}
     \begin{subfigure}
         \centering
         \includegraphics[width=1.6in]{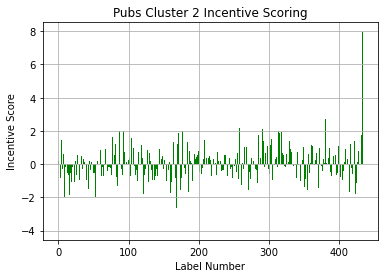}
         \end{subfigure}
     \begin{subfigure}
         \centering
         \includegraphics[width=1.6in]{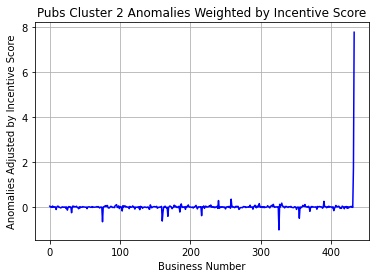}
         \end{subfigure} 
        \caption{Pubs label violations, incentive score, and weighted incentive score.}
        
        \label{publabelviolations}
\end{figure*}

\begin{figure*}
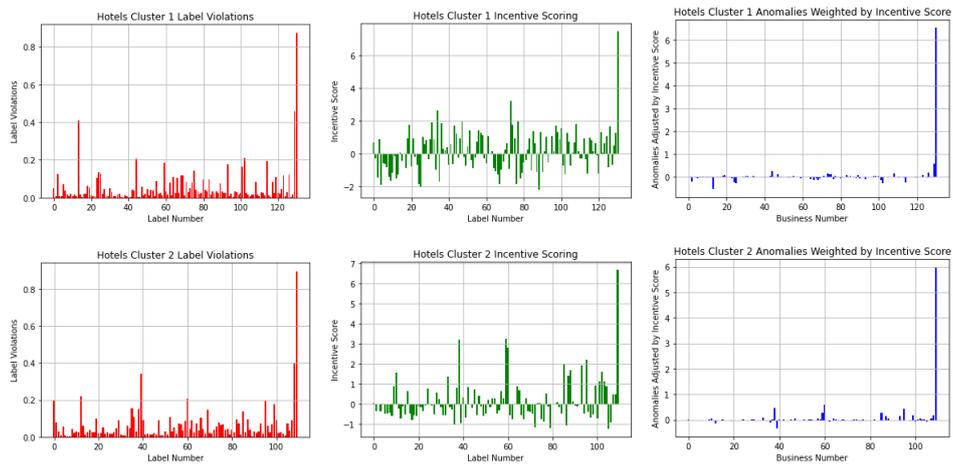

     \centering
     \begin{subfigure}
         \centering
         \includegraphics[width=1.6in]{hotelcluster1violations.png}
      \end{subfigure}
     \begin{subfigure}
         \centering
         \includegraphics[width=1.6in]{hotelcluster1incentive.png}
         \end{subfigure}
     \begin{subfigure}
         \centering
         \includegraphics[width=1.6in]{hotelcluster1weighted.png}
         \end{subfigure}

     \begin{subfigure}
         \centering
         \includegraphics[width=1.6in]{hotelcluster2violations.png}
      \end{subfigure}
     \begin{subfigure}
         \centering
         \includegraphics[width=1.6in]{hotelcluster2incentive.png}
         \end{subfigure}
     \begin{subfigure}
         \centering
         \includegraphics[width=1.6in]{hotelcluster2weighted.png}
         \end{subfigure} 
        \caption{Cinema label violations, incentive score, and weighted incentive score.}
\end{figure*}

\begin{figure*}
     \centering
     \begin{subfigure}
         \centering
         \includegraphics[width=1.6in]{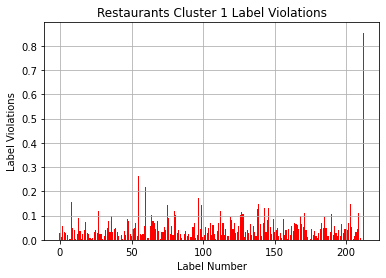}
      \end{subfigure}
     \begin{subfigure}
         \centering
         \includegraphics[width=1.6in]{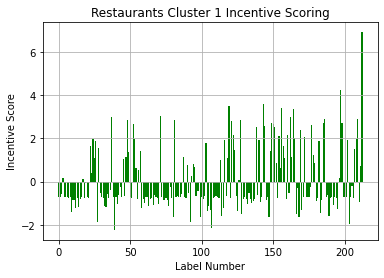}
         \end{subfigure}
     \begin{subfigure}
         \centering
         \includegraphics[width=1.6in]{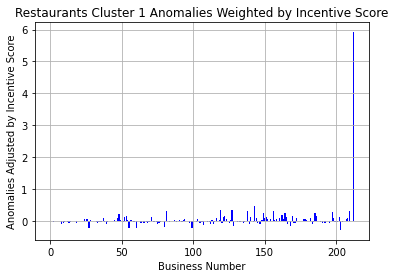}
         \end{subfigure}

     \begin{subfigure}
         \centering
         \includegraphics[width=1.6in]{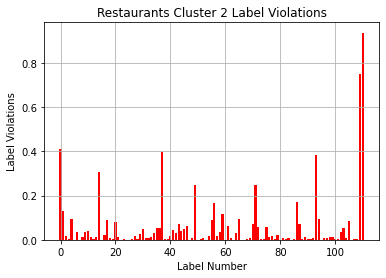}
      \end{subfigure}
     \begin{subfigure}
         \centering
         \includegraphics[width=1.6in]{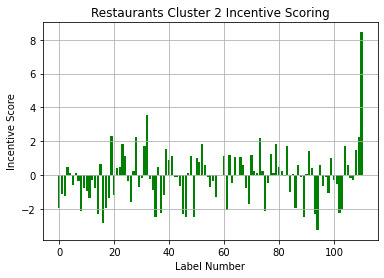}
         \end{subfigure}
     \begin{subfigure}
         \centering
         \includegraphics[width=1.6in]{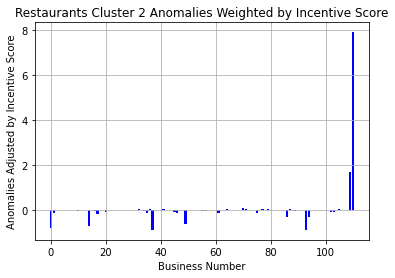}
         \end{subfigure} 
        \caption{Restaurants label violations, incentive score, and weighted incentive score.}
        \label{restaurantsviolations}
        
\end{figure*}

\begin{figure*}
     \centering
     \begin{subfigure}
         \centering
         \includegraphics[width=1.6in]{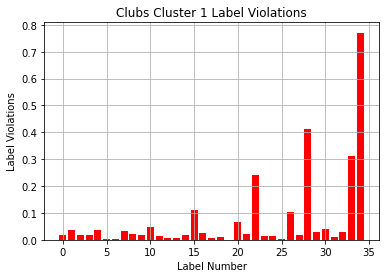}
      \end{subfigure}
     \begin{subfigure}
         \centering
         \includegraphics[width=1.6in]{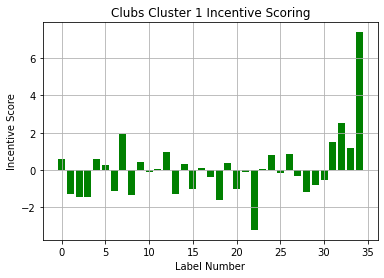}
         \end{subfigure}
     \begin{subfigure}
         \centering
         \includegraphics[width=1.6in]{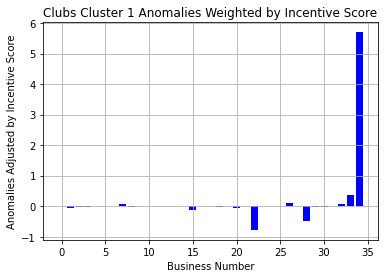}
         \end{subfigure}

     \begin{subfigure}
         \centering
         \includegraphics[width=1.6in]{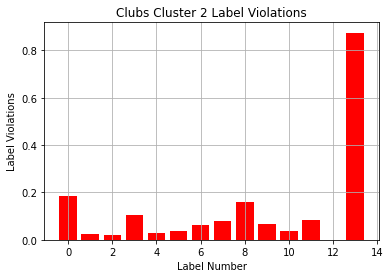}
      \end{subfigure}
     \begin{subfigure}
         \centering
         \includegraphics[width=1.6in]{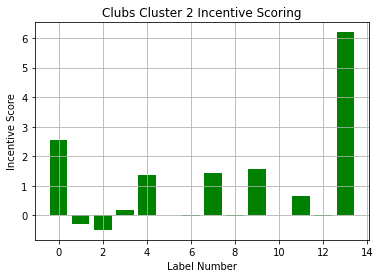}
         \end{subfigure}
     \begin{subfigure}
         \centering
         \includegraphics[width=1.6in]{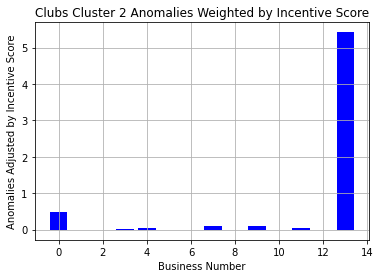}
         \end{subfigure} 
        \caption{Clubs label violations, incentive score, and weighted incentive score.}
\end{figure*}

\subsection{Weighted Spot Price Calculation}
\begin{figure}[t]
\centering
\includegraphics[width=3in]{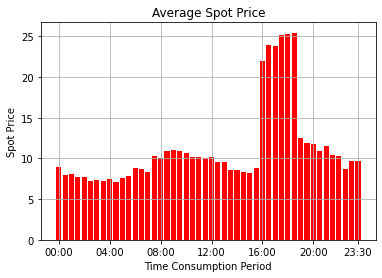}
\caption{Average spot price for each consumption period.}
\label{averagespot}
\end{figure}

For the weighted spot price calculation, we used data provided by Octopus energy prices. As model building relies on using summer data, we employ equivalent summer data to build our weighted spot price. The Octopus data is split into 14 sub-regions accounting for regions within the UK, such as London, East England, and Midlands. We note that despite these regions being split into groups, the level of inter-regional price volatility is low. For simplicity, we take a simple mean average of all these regions combined, which is used as the basis for our weighted spot price. This average spot price per consumption period is shown in Figure \ref{averagespot}, and we show the corresponding incentive weighting array in Figure \ref{incentivescore}.


      

\begin{figure}[t]
\centering
\includegraphics[width=3in]{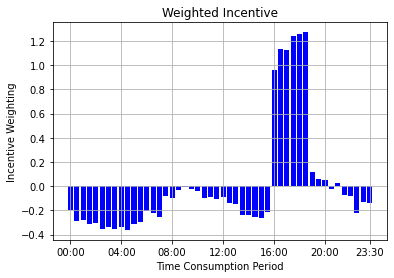}
\caption{Incentive weighting for each consumption period.}
\label{IW}
\end{figure}

\begin{figure}[t]
\centering
\includegraphics[width=3in]{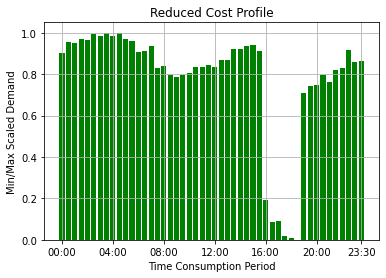}
\caption{Reduced cost red-team profile for each consumption period.}
\label{reducedcostprofile}
\end{figure}

\subsection{False Data Injection Profiles}
We introduce two FDI style profiles into the load datasets. One profile is a simple meter bypass, which is typical of the current state of play in physical attacks; in meter bypass, the profile is simply replaced with a set of 0s representing no load. We also introduce a more sophisticated reduced-cost spot attack (RCSA) profile. This RCSA profile represents an attacker attempting to create a non-zero load profile to attain significant reductions via the spot price. The RCSA profile is shown in Figure \ref{reducedcostprofile}

\subsection{Industrial Cluster Groups}
Figure \ref{clustergroupings} shows the individual measurement sets for the cluster groupings, while Figure \ref{averageclustergroupings} presents the average corresponding model. These models are built using the 2009 summer data set. We note that although there are various industrial business models, we often see a trend towards a limited number of common models not dissimilar to the typical consumer load. We also illustrate this in \ref{Histograms} which shows the time dynamics and relative density of the respective profiles in 3d.

In Cinema cluster 1, we observe a consumption peak at approximately 12:00, which falls off after around 16:00. This shape is somewhat unusual as we might expect typical a cinema business to continue into the late evening. However, it may be necessary to distinguish between 'mom-and-pop' style cinemas, which might shut relatively early, and large cinema corporations that run into the night. Cinema 2 resembles a more typical consumer load flow profile, with a broad consumption peak running from 08:00 to 20:00. 

The most atypical business grouping in the dataset is Club cluster 1. Where the other businesses have the expected peaks around the common spot peak consumption times, Club cluster 1 exhibits a later peak at around the 20:00 mark, which gradually increases into the night. This is consistent with the nature of business, given that the main hours of operation for nightclubs are during the night. 

Hotel cluster 1 exhibits a notable peak at 06:00, which is potentially attributable to breakfast preparations. Curiously, a similar peak does not occur for lunch, but we do see it for the dinner menu at approximately 16:00. Clubs cluster 2 is also atypical in that it has a low range between the midday peaks and the overnight operation. Similar to other groups Hotels cluster 2 and Pubs cluster 1 exhibit similar patterns to a residential consumer load profile. However, we note that, generally, industrial profiles have broader operation periods, which is reflected in the peak and intraday consumption windows.

\subsection{False Profile Detection}
The performance of the detection model is shown in Figures \ref{hotelviolations} to \ref{restaurantsviolations}, indicating the violation score, incentive weight, and incentive-weighted violation score for both identified hotel cluster types. The initial cluster models for each respective business were built using the 2009 summer weekend dataset through a combined clustering and scoring approach. They are then cross-compared with the 2010 summer weekend dataset using the anomaly detection technique. In each of these figures, the last two data points are the bypass vector and RCSA attack, respectively. Consistent detection levels were observed for the RCSA attack for all three types of detection. Incentive scoring adds a heavy weighting to detection for the RCSA attack. For both clusters 1 \& 2, incentive-weighted detection for the RCSA attack was clear and present. We note a similar result in Figures \ref{publabelviolations} to \ref{restaurantsviolations}, with consistent incentive-weighted detection for the RCSA style of attack. However, note that incentive-weighted detection is not as effective for the bypass attack. The bypassed data is not easily identified using the incentive-weighted approach. Indeed, we do see generally higher than average scoring for the bypass vector in some cluster groups (e.g., clubs cluster 1). Generally, however, this method of scoring for this type of attack is undermined due to incentive weighting. As the bypass has no clear incentive weighting, this reduces the impact.

\section{Conclusion \& Future Work}
Smart meters are a weakly defended, distributed infrastructure that represent an easy attacking opportunity for a cyber attacker. Altering the load profile in a smart meters can provide financial incentives opportunities for attackers with few current opportunities to detect this threat.  

In this paper, we have examined an incentive weighted detection model for FDI style attacks against load-profile datasets. Through feature-based clustering, we examined different groupings within industrial load profiles and created an incentive-weighted detection methodology to examine potential fraud. In short, this work has investigated how to improve corporate fraud detection in smart data through clustering and an incentive-weighted detection approach. 

In the first contribution of this paper, we examined how to establish a combination of hierarchical clustering and silhouette scoring base models for industrial load profiles. We incorporated real-life datasets and, for illustrative purposes, analysed businesses from entertainment-style industrial parks, due to the general familiarity of the nature of these businesses.

We injected fraudulent profiles into the datasets. These were based on two likely red-team cases: first, a bypass profile representing a common `direct bypass’ of the metering infrastructure; and second, an RCSA profile, which attempts to exploit variations in price resulting from spot pricing.

The previously outlined base models were then used to enhance the detection of the injected profiles using an incentive-weighted violation approach. This approach incorporates both spot pricing and the level of departure from the expected model.

In future work, dis-aggregation of the modern additions to the distribution network which influence load profiles which be a useful 'future proofing' of the model for contemporary power systems. These could include solar panels, heat pumps, and storage devices. It would be worthwhile to understand how these might impact the detection of attacks against load profiles.

\section{Acknowledgements}
This work was performed as part of the Analytical Middleware for Informed Distribution Networks (AMIDiNe) project under EPSRC grant EP/S030131/1.

\bibliographystyle{iet.bst}
\bibliography{powerpaper.bbl}

\end{document}